\documentclass[12pt]{article}

\usepackage{amsmath,amssymb,graphicx,amsthm, amssymb,sidecap}

\begin{document}

\centerline{\textbf{\Large The odd couple: quasars and black holes}}

\bigskip

\centerline{Scott Tremaine}
\centerline{Institute for Advanced Study}
\centerline{Princeton, NJ 08540, USA}

\bigskip

\paragraph{Abstract} {\small Quasars emit more energy than any other objects in the universe, yet are not much bigger than the solar system. We are almost certain that quasars are powered by giant black holes of up to $10^{10}$ times the mass of the Sun, and that black holes of $10^6$ to $10^{10}$ solar masses---dead quasars---are present at the centers of most galaxies. Our own galaxy contains a black hole of 4.3 million solar masses. The mass of the central black hole appears to be closely related to other properties of its host galaxy, such as the total mass in stars, but the origin of this relation and the role that black holes play in the formation of galaxies are still mysteries. }

\bigskip

\bigskip

\noindent
Black holes are among the most alien predictions of Einstein's general theory of relativity: regions of spacetime in which gravity is so strong that nothing, not even light, can escape. More precisely, a black hole is a singularity in spacetime surrounded by an event horizon, a surface that acts as a perfect one-way membrane: matter and radiation can enter the event horizon, but once inside can never escape. Although black holes are an inevitable consequence of Einstein's theory, their main properties were only understood---indeed, the name was only coined---a half-century after Einstein's work \cite{chandra}. Remarkably, an isolated, uncharged black hole is completely characterized by only two parameters, its mass and angular momentum. 

An eloquent tribute to the austere beauty of these objects is given by the astrophysicist Subrahmanyan Chandrasekhar in the prologue to his monograph {\it The Mathematical Theory of Black Holes}\,: ``The black holes of nature are the most perfect macroscopic objects there are in the universe: the only elements in their construction are our concepts of space and time. And since the general theory of relativity provides only a single unique family of solutions for their descriptions, they are the simplest objects as well''---although anyone who scans the six hundred pages that follow is unlikely to agree that they are as simple as claimed! Simple or complex, black holes have been studied by mathematicians, physicists, and astronomers for the past five decades and we are now almost certain, for reasons outlined in this essay, that black holes do exist and that a giant black hole of several million times the mass of the Sun is present at the center of our galaxy.

Laboratory study of a black hole is impossible with current or foreseeable technology, so the only way to test these predictions of Einstein's theory is to find black holes in the heavens. Not surprisingly, isolated black holes are difficult to see. Not only are they black, they are also very small: a black hole with the mass of the Sun is only a few kilometers in diameter (this statement is deliberately vague since there is no unique notion of ``distance'' close to a black hole). However, the prospect for detecting black holes in gas-rich environments is much better: the gas close to the black hole will normally take the form of a rotating disk, called an accretion disk since the orbiting gas gradually spirals in towards the event horizon as it loses orbital energy, most likely due to magnetohydrodynamic turbulence in the disk \cite{balbus}. The orbital energy is transformed into thermal energy, which heats the disk gas until it begins to glow, mostly at ultraviolet and X-ray wavelengths. By the time the inspiraling gas disappears behind the event horizon, deep within the gravitational well of the black hole, a vast amount of radiation has been emitted from every kilogram of accreted gas.

In this process, the black hole can be thought of as a furnace: when provided with fuel (the inspiraling gas) it produces energy (the outgoing radiation). Einstein's iconic formula $E=Mc^2$ relates mass $M$ and the speed of light $c$ to an energy $E$ called the rest-mass energy. Using this relation there is a natural dimensionless measure of the efficiency of this or any other furnace: the ratio of the energy it produces to the rest-mass energy of the fuel that it consumes. For furnaces that burn fossil fuels the efficiency is extraordinarily small, about $5\times10^{-10}$, and all combustion processes based on chemical reactions have similarly low efficiencies. For fission reactors using uranium fuel the efficiency is much better, around 0.1\%; and for the fusion reactions that power the Sun and stars the efficiency can reach 0.3\%.  Black-hole furnaces can have far higher efficiency than any of these, between 10\% and 40\% for accretion of gas from a thin accretion disk. In the unlikely event that we could ever domesticate black holes, the entire electrical energy consumption in the U.S. could be provided by a black-hole furnace consuming only a few kilograms of fuel per year. 

\bigskip

\noindent
Despite the relatively low efficiency of fusion reactions, most of the light in the universe comes from stars. Most of the stars in the universe are organized in galaxies---assemblies of up to $10^{11}$ stars orbiting in a complex dance determined by their mutual gravitational attraction. Our own galaxy contains a few times $10^{10}$ stars arranged in a disk; the nearest of these is about 1 parsec (3.26 light years) from us, and the distance to the center of our galaxy is about 8 kiloparsecs. The diffuse light from distant stars in the galactic disk comprises the Milky Way \cite{binney}.

A small fraction of galaxies contain mysterious bright, compact sources of radiation at or near their centers, called active galactic nuclei \cite{krolik}. The brightest of these are the quasars; remarkably, these can emit up to $10^{13}$ times the luminosity of stars like the Sun, thereby outshining the entire galaxy that hosts them. Even though quasars are much rarer than galaxies, they are so bright that they contribute almost 10\% of the light emitted in the universe. 

Ironically, the extraordinary luminosity of quasars is what made them hard to discover: except in a few cases, they are so bright that the host galaxy cannot be seen in the glare from the quasar, and so small that they appear as point sources even at the angular resolution of the Hubble Telescope, so they look just like stars (in fact ``quasar" is a contraction of ``quasi-stellar object"). Thus even the brightest quasars are usually indistinguishable from millions of stars of similar brightness. Fortunately, some quasars are also strong sources of radio emission, and in 1963 this clue enabled Maarten Schmidt at Caltech to identify a radio source called 3C 273 with a faint optical source that otherwise looked like an undistinguished star \cite{schmidt}. With this identification in hand Schmidt was able to show that 3C 273's spectral lines were redshifted---Doppler shifted to wavelengths 16\% longer than laboratory spectra by the cosmological expansion of the universe---and thus that 3C 273 was at a distance of 800 megaparsecs, ten million times further away than it would have been if it were a normal star. 

By now almost 100,000 quasars are known. The most distant of them is almost 100 times as far away as 3C 273 and its light was emitted when the universe was only 6\% of its present age. However, the formation of quasars this early in the history of the universe was a very rare event. Most were formed when the universe was 20--30\% of its current age, and quasars today are a threatened species: the population has declined from its peak by almost two orders of magnitude, presumably because the fuel supply for quasars dried up as the universe expanded at an accelerating rate. 

\bigskip

\noindent
How can quasars emit so much energy? The suggestion that they are black-hole furnaces was made soon after they were discovered, independently by Edwin Salpeter in the U.S. and Yakov Zel'dovich in the Soviet Union. However, in the 1960's the black hole was a novel and exotic concept, and the black-hole masses required to explain quasar properties were staggering, roughly $10^8$ times the solar mass. Thus most astronomers quite properly focused on more conservative models for quasars, such as supermassive stars, dense clusters of ordinary stars or neutron stars, collapsing gas clouds, etc. Over the next two decades all of these models were subjected to intense scrutiny, which revealed more and more difficulties in matching the growing body of observations of quasars; meanwhile other studies showed that such systems generically evolve into a single black hole containing most of the mass of the original system, thereby suggesting that the formation of massive black holes was natural and perhaps even inevitable. 

A number of indirect but compelling arguments also support the black-hole furnace hypothesis. 

The first of these relates to efficiency. The luminous output of a bright quasar over its lifetime corresponds to a rest-mass energy of about $10^8$ times the mass of the Sun. If this were produced by the fusion reactions that power stars, with the efficiency of 0.3\% given earlier, the total mass of fuel required would be $3\times10^{10}$ solar masses---almost the total of all the stars in our galaxy. There is no plausible way to funnel this much mass into the tiny central region that the quasar occupies, and no evidence that so much mass resides there. On the other hand, for a black-hole furnace the efficiency is 10\% or more, so the required mass is less than $10^9$ solar masses, and this much gas is not hard to find close to the center of many galaxies. Thus the black-hole furnace is the only model that does not bankrupt the host galaxy's fuel budget. 

A second argument is based on the small size of quasars. We have known since their discovery that quasars appear as point sources even in the best optical telescopes, and this observation alone implies that their sizes must be less than about a kiloparsec. Using long-baseline radio interferometry the upper limit to the sizes of some quasars can be reduced to about one parsec. Even stronger limits on the size come from other measurements. Quasars vary irregularly in brightness on a wide range of timescales from weeks to decades and probably even longer. It proves to be quite difficult to construct any plausible model of a luminous astrophysical object that varies strongly on a timescale smaller than the time it takes light to travel across the object---the separate parts of the object are not causally connected on this timescale, so they vary independently and their contributions tend to average out. This argument suggests that the size of the most rapidly varying quasars must be less than the distance light travels in a few weeks, around a few hundredths of a parsec or a few thousand times the Earth-Sun distance. This upper limit is consistent with size estimates from a number of other methods, such as reverberation mapping, photoionization models, and gravitational lensing \cite{blr}. A few hundredths of a parsec is large by our standards but extremely small on galactic scales, a millionth of the size of the galaxy as a whole. A $10^8$ solar mass black hole and its surrounding accretion disk would fit comfortably inside this volume---its relativistic event horizon has a radius of about the Earth-Sun distance---but almost all of the alternative models that might explain quasars fail to do so. 

In a few cases space-based observatories can measure X-ray spectral lines emitted by the quasar. These are not the narrow lines seen in spectra of the Sun, stars, or interstellar gas; instead they are grossly misshapen, with broad tails extending to much longer wavelengths than the line would have in the lab. The only plausible explanation for these distortions is that they arise from gravitational redshift---the loss of energy as the X-rays climb out of a deep gravitational well on their way to us---and/or extreme Doppler shifts arising from relativistic motions, most probably in a rapidly rotating accretion disk. Either explanation requires that the X-rays were emitted from a region only a few times larger than the event horizon of a black hole, as no other known astrophysical system has such high velocities and deep potential wells \cite{mil}. 

Some quasars emit powerful jets of plasma (Figure 1) that extend for up to a megaparsec \cite{jet}. The production of these jets is not so remarkable---for example, various kinds of star also produce jets, although on a much smaller scale. What is more striking is that quasar jets typically travel at close to the speed of light. Once again, there is no plausible way to produce such high velocities except close to the event horizon of a black hole. Moreover in most cases the jets are accurately straight, even though the innermost plasma in the jet was emitted a million years after the material at the far end. Thus whatever mechanism collimated the jet must maintain its alignment over million-year timescales; this is easy to do if the jets are squirted out along the polar axis of a spinning black hole, but difficult or impossible in other quasar models.

Finally, there is strong evidence that a handful of systems that emit strong X-ray radiation consist of a normal star and a black hole. These black holes are much less massive than those in quasars, only a few times the mass of the Sun \cite{mcc}. The black hole and the normal star orbit one another at a small enough distance---a few stellar radii---that material lost from the normal star fuels a miniature black-hole furnace. These systems reinforce our confidence in  the existence of black holes, and allow us to refine our understanding of the complex physics of a black-hole furnace.  

\begin{figure}
\centerline{\includegraphics[width=\textwidth]{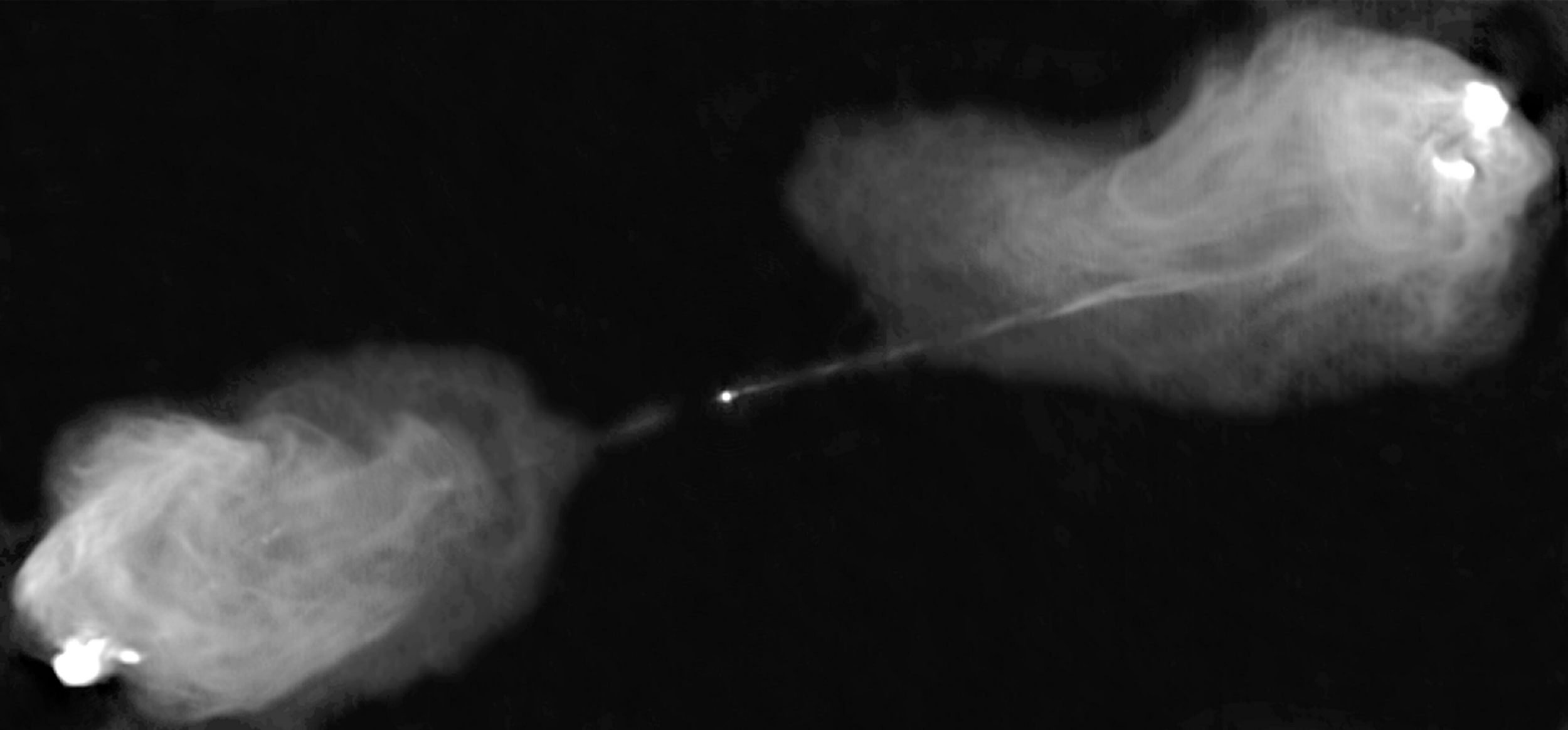}}
\caption{\small An image at radio wavelengths of jets from  the quasar Cygnus A.   The unresolved bright spot at the center of the image is the quasar, which is located in a galaxy 240 megaparsecs away. The long, thin jets emanating from the quasar terminate in bright ``hot spots''  when they impact the intergalactic gas that surrounds the galaxy. The hotspots are roughly 70 kiloparsecs from the quasar. The brighter of the two jets is traveling toward us; its brightness is boosted by relativistic effects. Image courtesy of NRAO/AUI; reproduced by permission of the AAS. From  R. A. Perley et al., ``The jet and filaments in Cygnus A'', Astrophysical Journal (1984): 285, L35-L38.}
\end{figure}

\bigskip

\noindent
Based on these and other arguments, almost all astrophysicists are persuaded that the power source for quasars is accretion onto massive black holes. Accepting this position leads to a simple syllogism that has driven much of research on this subject for the past several decades: if quasars are found in galaxies, and the number of quasars shining now is far smaller than when the universe was young, and quasars are black-hole furnaces, then many ``normal'' galaxies should still contain the massive black holes that used to power quasars at their centers, but are now dark. Can we then find these ``dead quasars'' in nearby galaxies?

There are two important guideposts in the search for dead quasars. The first came from a simple argument by a Polish astronomer, Andrzej So\l tan \cite{sol}. We know that the universe is homogeneous on large scales, and therefore on average  the energy density in quasar light must be the same everywhere in the universe (here ``average'' means averaged over scales greater than about 10--20 megaparsecs, which is still small compared to the overall ``size'' of the universe, a few thousand megaparsecs). We can measure this energy density by adding up the contributions from all the quasars found in surveys, after straightforward corrections for incompleteness. If this energy were produced by black-hole furnaces with an efficiency of 10\% (say), then a mass $M$ of material accreted by black holes would produce $0.1Mc^2$ in quasar light. Similarly, if the average mass density of dead quasars is $\rho$, then the energy density of quasar light must be $0.1\rho c^2$. Since we know the latter figure we can invert the argument to determine the mass density of dead quasars. The power of this argument is that it requires no assumptions about the masses or numbers of black holes, no knowledge of when, where, or how quasars formed, and no understanding of the physics of the quasar furnace except its efficiency. So\l tan's argument tells us that the density of dead quasars should be a few times $10^5$ solar masses per cubic megaparsec, compared to a density of large galaxies of about one per hundred cubic megaparsecs. What it does not tell us is how common they are: on average there could be one dead quasar of $10^7$ solar masses in every galaxy, or one of $10^9$ solar masses in 1\% of galaxies.

The second guidepost is that the centers of galaxies are the best places to prospect for dead quasars. There are several reasons for this. First, live quasars seem to be found near the centers of their host galaxies, although this is difficult to tell with precision because the glare from the quasar obscures the structure of the host. Second, the fuel supply for a black hole sitting at rest in the center of the galaxy is likely to be much larger than for one orbiting in the outskirts of the galaxy.  Third, massive black holes orbiting in a galaxy tend to lose orbital energy through gravitational interactions with passing stars, so they spiral in to the center of the galaxy \cite{df}. And finally, like the drunkard looking for his keys under the lamppost, we look for dead quasars at the centers of galaxies because that is where it's easiest to find them: the search area is small and the density of stars that are affected by the black hole's gravitational field is high. 

Stars that come under the influence of the black hole's gravitational field---typically those within a distance of a few tenths of a parsec to a few tens of parsecs, depending on the black hole mass---are accelerated to higher velocities; although individual stars cannot be detected in galaxies other than our own, this speedup leads to increased Doppler shifts which broaden the spectral lines from the collective stellar population. This broadening can be detected by spectroscopic observations with sufficiently high spatial resolution and signal-to-noise ratio. The search for this effect in the centers of nearby galaxies began around 1980, and yielded evidence for black holes in a handful of nearby galaxies. Strictly, the evidence was for dark objects with masses of millions to billions of solar masses, as the spatial resolution of these observations was still $10^5$ times the size of the event horizon of the putative black hole. These results were tantalizing, but incomplete: the problem was that the angular resolution of ground-based telescopes is limited by blurring caused by the atmosphere, so the effects of a black hole could be detected only in the closest galaxies, and then only over a limited range of distances from the center.  Precisely this problem was one of the major motivations for constructing the Hubble Space Telescope, which at the time of its launch in 1990 had roughly ten times the angular resolution of the best ground-based telescopes. Since then the Hubble Telescope has devoted many hundreds of hours to the hunt for black holes at the centers of galaxies, and by now Hubble has confirmed and strengthened the ground-based detections in nearby galaxies and produced firm evidence for black holes in over two dozen more distant ones \cite{gul}. Even in the best cases, this method can only probe to a few tenths of a parsec from the galaxy center, but we are persuaded that the massive dark objects observed by Hubble must be black holes because the alternatives (e.g., a cluster of low-luminosity stars) are far less plausible. By now the Hubble Telescope has turned to other tasks, but the search for dead quasars has been resumed  by ground-based telescopes, now using adaptive optics systems that can correct for atmospheric blurring in real time. Adaptive optics is beginning to provide angular resolutions that equal or exceed Hubble's as well as far higher signal-to-noise ratios, since the collecting area of the biggest ground telescopes is ten times that of Hubble.

The painstaking measurement of stellar motions near the centers of galaxies has been supplemented by an unexpected gift from the heavens: the otherwise unremarkable galaxy NGC 4258 contains at its center a thin, nearly flat, rotating disk of gas, about a tenth of a parsec in radius. The gas includes water vapor, and the temperature and density in the disk are right for the production of maser (microwave laser) emission in the water, stimulated by a weak active galactic nucleus at the center of the disk. The maser emission consists of tiny, intensely bright sources of radiation in the spectral lines of water, and by measuring the Doppler shift of these sources and their motion across the sky using an intercontinental array of radio telescopes we can map out the rotation of the disk with exquisite precision. The disk is found to rotate around the active nucleus; from the disk kinematics we can deduce that the nucleus is much smaller than the inner radius of the disk, and that its mass is $3.78\times10^7$ solar masses with a measurement uncertainty of only 0.3\% (possible systematic errors due to the choice of model are bigger, about 1\%) \cite{her}---the strongest evidence we have for a massive dark object at the center of any distant galaxy. 

Finally, our own galaxy offers unique evidence for a black hole at its center \cite{gen}. Very close to the geometric center of the distribution of stars in the Milky Way is a compact source of strong radio emission, known as Sagittarius A*.  This region is difficult to study because small solid particles in interstellar space (commonly called ``dust'' but more like smoke) obscure visible light coming from stars near the center. The smoke can be penetrated by infrared radiation, and at these wavelengths high-resolution observations reveal a handful of bright stars within a few hundredths of a parsec from Sagittarius A*. The positions and velocities of these stars have been tracked, some for as long as two decades; in particular, the star S2 has an orbital period of only 15.8 years and now has been tracked through more than one complete orbit. Some four centuries ago, Johannes Kepler showed that the orbits of the planets around the Sun were ellipses with the Sun at one focus; here the orbit of S2 is also an ellipse, with Sagittarius A* at one focus (Figure 2). Using freshman mechanics, we can deduce from this orbit that the body located at this focus has a mass of $4.3\times10^6$ solar masses, with an uncertainty of less than 10\%, and that the size of this body is less than only 100 times the Earth-Sun distance or a few thousand times the radius of the event horizon for a black hole of this mass.  This extreme concentration of mass is incompatible with any known long-lived astrophysical system other than a black hole. The center of our galaxy thus offers the single best case for the existence of black holes and strongly suggests that the massive dark objects found in the centers of other galaxies are also black holes. 

\begin{figure}
\vspace{-1.3in}
\centering
\includegraphics[width=0.8\textwidth]{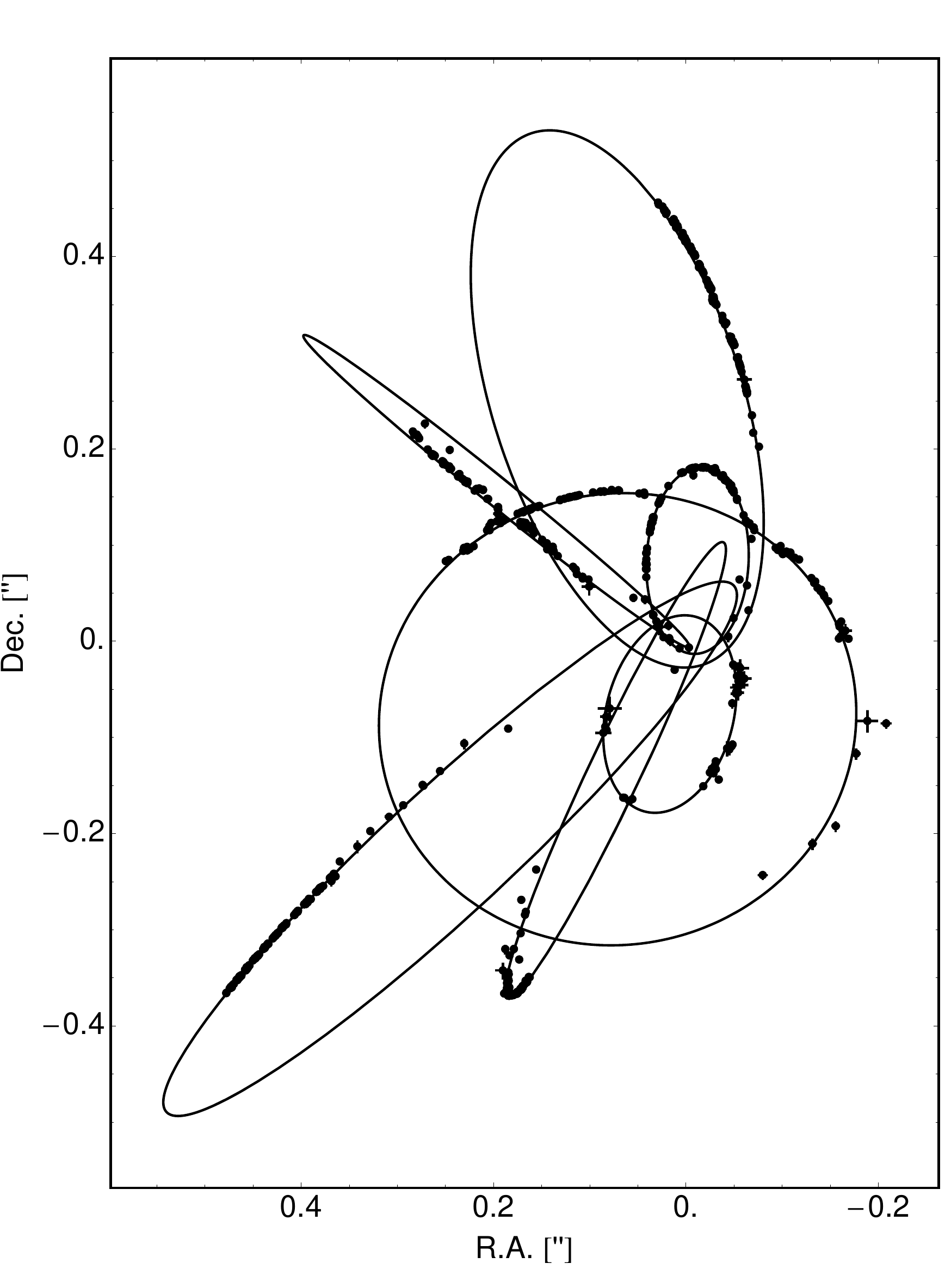}
\caption{\small Orbits of the stars near the center of our galaxy. The radio source Sagittarius A*, believed to coincide with the black hole at the galaxy center, is at the origin of the coordinates. The width of the frame is 0.03 parsecs or 6700 times the Earth-Sun distance. Each orbit is well-fit by an ellipse with one focus at Sagittarius A* (the focus is not on the symmetry axis of the orbit because the normal to the orbit is inclined to the line of sight). The star on the smallest orbit, called S2, has a period of 15.8 years, and its point of closest approach to Sagittarius A* is 120 times the Earth-Sun distance. These parameters imply that Sagittarius A* is associated with a mass of $4.3\times10^6$ solar masses contained within a radius of about 100 times the Earth-Sun distance. Courtesy of R.\ Genzel and S.\ Gillessen.}
\end{figure}

What else have we learned from these discoveries? First, black holes seem to be present in most galaxies, except perhaps for a class known as late-type galaxies. Second, the mass of the black hole is strongly correlated with the mass or luminosity of the galaxy; roughly, the black-hole mass is about 0.2\% of the mass of the stars in the galaxy. A final question is whether the black holes we are finding in nearby galaxies are really dead quasars. From galaxy surveys we can determine the average mass density in stars in the local universe, and since black-hole masses are typically 0.2\% of the stellar mass in a galaxy, we can estimate the mass density of black holes in the local universe. On the other hand So\l tan's argument, described earlier, gives the expected mass density of dead quasars in the local universe from completely different data (surveys of distant quasars as opposed to surveys of nearby galaxies). The two estimates agree to within a factor of two or so, well within the uncertainties, so there is little doubt that the black holes we have found are indeed the ash from quasars or other active galactic nuclei. 

\bigskip

\noindent
This essay has described briefly what we have learned about the intimate relation between quasars, one of the most remarkable components of the extragalactic universe, and black holes, one of the most exotic predictions of twentieth-century theoretical physics. Many aspects of this relation remain poorly understood, and to close I will mention two of the most profound unanswered questions. 

The first of these is the relation between black holes and galaxy formation. Although black holes comprise only a fraction of a percent of the mass of the stars in galaxies, the energy released in forming them is hundreds of times larger than the energy released in forming the rest of the galaxy. If even a small fraction of the energy emitted by the black-hole furnace is fed back to the surrounding gas and stars, it would have a dramatic influence on the galaxy formation process. In an extreme case the quasar could blow the gas out of the galaxy and thereby quench the formation of new stars. Are black holes and quasars an interesting by-product of galaxy formation that has no influence on the formation process, or do they play a central role in regulating it? More succinctly, do galaxies determine the properties of quasars or vice versa?

The second profound question one of physics rather than astronomy. All of the tests of Einstein's theory so far---which it has passed with flying colors---have been conducted in weak gravitational fields, such as those on Earth or in the solar system. In contrast, we have no direct evidence that the theory works well in strong gravitational fields. Many naturally occurring processes near black holes in galaxy centers---tidal disruption of stars, swallowing of stars, accretion disks, and even black-hole mergers---can potentially be measured with the next generation of astronomical observatories. Can we understand these processes well enough to test the unique predictions of general relativity for physics in strong gravitational fields, and will Einstein turn out to be right?

\end{document}